\documentclass{article}

\usepackage[preprint]{neurips_2026}
\usepackage[utf8]{inputenc} 
\usepackage[T1]{fontenc}    

\usepackage{booktabs}       
\usepackage{amsfonts}       
\usepackage{nicefrac}       
\usepackage{microtype}      
\usepackage[table]{xcolor}
\usepackage{amsmath} 
\usepackage{graphicx}
\usepackage{float}

\usepackage{hyperref}       
\usepackage{url}            

\definecolor{sectionpink}{RGB}{245,210,210}
\definecolor{selfblue}{RGB}{217,235,245}

\usepackage{pgfplots}
\pgfplotsset{compat=1.18}
\usetikzlibrary{calc}
\usepgfplotslibrary{groupplots}
\usepgfplotslibrary{fillbetween}
\usepackage{wrapfig}
\usepackage{multirow}
\usepackage{enumitem}

\usepackage{algorithm}
\usepackage{algpseudocode}

\usepackage{xspace}
\newcommand{\method}{PreLort\xspace}
\title{PreLort: Prefix-Nested LoRA for Federated Fine-Tuning under Rank Heterogeneity}

\author{
\textbf{Muhammad Waseem$^{1}$,
Nurbek Tastan$^{1}$,
Andrej Jovanovic$^{2,3}$,
Nicholas D. Lane$^{2,3}$} \\
\textbf{Nils Lukas$^{1}$,
Karthik Nandakumar$^{1,4}$,
Samuel Horv\'ath$^{1}$} \\
$^{1}$MBZUAI, UAE 
\hspace{0.6cm} $^{2}$University of Cambridge, UK \\
$^{3}$Flower Labs, UK 
\hspace{0.3cm} $^{4}$Michigan State University, USA
}
\begin{document}

\maketitle

\begin{abstract}
    Federated fine-tuning of large language models using parameter-efficient methods such as LoRA enables privacy-preserving adaptation of foundation models. Heterogeneous hardware resources introduce challenges, as clients with different adapter ranks cannot be directly aggregated. While existing methods enable aggregation under heterogeneous ranks, they fail to control how information is distributed across rank dimensions, leading to suboptimal use of shared low-rank representations. Instead, we propose \textbf{PreLort}: a nested low-rank formulation for federated LoRA that organizes adapter dimensions into a prefix hierarchy. Our approach ensures that lower-rank dimensions encode task-relevant information, while higher-rank dimensions capture additional capacity. Building on this, we introduce (i) a segment-wise aggregation rule that averages only over clients contributing to each rank segment, avoiding dilution from zero-padded lower-rank clients, and (ii) a prefix-nested training strategy that optimizes each adapter under multiple rank truncations, encouraging useful signal to concentrate in low-rank prefix dimensions. Together, these components encourage a consistent low-rank prefix capturing the most task-relevant information, while higher-rank dimensions learn additional capacity. This allows low-rank clients to benefit from richer information contributed by higher-rank clients, as prefix dimensions are consistently learned and aggregated. Experiments demonstrate that our method consistently outperforms prior heterogeneous federated LoRA methods in accuracy and ROUGE-L, while achieving lower or comparable perplexity across multiple base models. 
\end{abstract}

\section{Introduction}

Federated learning (FL) enables collaborative training without sharing raw data~\citep{pmlr-v54-mcmahan17a}, but it does not inherently guarantee privacy, as model updates can leak sensitive information. Stronger privacy guarantees typically require additional techniques such as differential privacy or secure aggregation~\citep{kairouz}. Large language models (LLMs) are increasingly adapted to downstream tasks via fine-tuning, but their scale makes full-parameter training impractical in decentralized environments~\citep{zhang2024buildingfederatedgptfederated}. To reduce computational and communication costs, parameter-efficient fine-tuning (PEFT) methods such as LoRA, prefix tuning, and prompt tuning have become the standard approach, learning low-rank updates while keeping the base model frozen \citep{hu2022lora, houlsby2019parameter, pfeiffer-etal-2021-adapterfusion, lester-etal-2021-power} .

A key challenge in federated LoRA is \emph{rank heterogeneity}~\citep{cho-etal-2024-heterogeneous, wang2024flora}. In realistic cross-device settings, clients operate under varying memory and compute constraints. As such, clients may initialize non-identical LoRA adapters, preventing direct aggregation of adapters. A native solution is to enforce a uniform rank across clients. This is suboptimal as enforcing minimal rank across all clients will produce a model that does not have enough representational power. Recent works have provided solutions to this problem enabling aggregation across heterogeneous LoRA modules through zero-padding \citep{wang2024flora, cho-etal-2024-heterogeneous, singhal-etal-2025-fedex}. However, these methods focus primarily on how to combine updates algebraically, without considering how information is distributed across rank dimensions.

This leads to a fundamental limitation: \emph{heterogeneous ranks induce misaligned representations}. Even when aggregation is mathematically well-defined, combining independently trained models may introduce inconsistencies due to misaligned representations or incompatible parameter updates~\citep{pmlr-v162-wortsman22a, ilharco2023editing, matena2022merging}. As a result, lower-rank clients do not benefit from higher-rank updates.

In this work, we argue that addressing rank heterogeneity requires not only improved aggregation, but also \emph{structured alignment of representations across ranks}. We introduce \textbf{PreLort}: a nested low-rank formulation that organizes LoRA adapter dimensions into a prefix structure, where lower-rank dimensions capture task-relevant information and higher dimensions provide additional capacity~\citep{kusupati2022matryoshka, cai2020once, yu2018slimmable,  Yu2019UniversallySN}. This structure encourages important information to concentrate hierarchically in prefix dimensions, enabling consistent aggregation across heterogeneous ranks.

Building on this formulation, we propose two key components. \textit{First}, we design a \emph{\textbf{nested training objective}} that explicitly trains multiple rank truncations within each client update, encouraging task-relevant signal to concentrate in shared low-rank prefixes hierarchically. This produces representations that remain useful under arbitrary truncation and are consistently aligned across clients. \textit{Second}, we introduce a \emph{\textbf{segment-wise aggregation rule}} that leverages this structure by aggregating adapter parameters only over the subset of clients that support each rank segment, avoiding zero-padding and preserving the contribution of higher-rank clients without introducing bias.

Our approach differs from prior work in that it jointly addresses representation alignment across rank dimensions, training dynamics, and aggregation under heterogeneous ranks. 

\vspace{0.1cm}

\textbf{Our main contributions are summarized as follows:}

\vspace{-0.2cm}

\begin{itemize}[leftmargin=0.4cm, itemsep=1pt]
    \item We identify representation misalignment across rank dimensions as a key limitation in heterogeneous federated LoRA.

    \item We propose a nested low-rank formulation that imposes a common prefix structure over adapter dimensions, and develop a unified training and aggregation strategy that enables consistent aggregation across heterogeneous ranks without zero-padding.

    \item We empirically demonstrate consistent improvements in model quality and stability over prior heterogeneous federated LoRA methods on instruction-tuning benchmarks.
\end{itemize}

\section{Related Work}

\paragraph{Parameter-Efficient Fine-Tuning.}
Parameter-efficient fine-tuning (PEFT) methods adapt large language models by training a small set of additional parameters while keeping the base model frozen. Among these, LoRA \citep{hu2022lora} represents weight updates as low-rank decompositions and has become a widely adopted approach for LLM adaptation. Other PEFT methods include adapter-based approaches \citep{houlsby2019parameter, pfeiffer-etal-2021-adapterfusion}, prompt-based methods such as prompt tuning and prefix tuning \citep{lester-etal-2021-power, li-liang-2021-prefix} and bias-only or partial finetuning techniques like BitFit \citep{ben-zaken-etal-2022-bitfit}. More recent work has explored improving efficiency and flexibility through adaptive rank allocation and structured updates, including LoFT~\citep{tastan2026loft}, AdaLoRA \citep{zhang2023adaptive}, DoRA \citep{liu2024dora}, and other variants that modify the parameterization or training dynamics of low-rank updates. While these approaches are effective in centralized settings, they do not directly address the challenges of heterogeneous resource constraints and rank variability that arise in federated learning.

\paragraph{Federated Fine-Tuning and Heterogeneous LoRA.}
Federated fine-tuning adapts large language models using decentralized data while avoiding direct data sharing. To reduce communication and computation costs, recent methods combine federated learning with parameter-efficient fine-tuning, especially LoRA. FedIT~\citep{zhang2024buildingfederatedgptfederated} applies LoRA with FedAvg for federated instruction tuning, but direct factor-wise averaging introduces aggregation noise because averaging LoRA factors does not correspond to averaging their induced updates. Recent work has also explored stronger privacy requirements in federated adaptation: BlindFed~\citep{tastan2025framework} considers a double-blind setting in which neither private client data nor the proprietary foundation model is directly exposed, using encrypted inference, low-rank parallel adapters, and secure aggregation. FLoRA~\citep{wang2024flora} addresses heterogeneous LoRA aggregation with stacking-based aggregation, which is exact in update space and naturally supports heterogeneous ranks. HetLoRA~\citep{cho-etal-2024-heterogeneous} further considers rank heterogeneity across clients through rank self-pruning and sparsity-weighted aggregation.

These methods enable heterogeneous LoRA aggregation, but they mainly focus on how to combine adapters after local training. In contrast, our work focuses on aligning the internal rank structure during training itself. By enforcing a prefix-nested organization of LoRA dimensions and aggregating rank segments only over contributing clients, \method encourages shared low-rank prefixes to capture task-relevant information while higher-rank dimensions provide additional capacity.

\paragraph{Slimmable Networks.}
Slimmable networks, introduced by \citep{Yu2019UniversallySN}, train a single model that can operate at multiple widths, enabling dynamic trade-offs between model capacity and efficiency. This idea has inspired follow-up work in federated learning, mainly for resource-aware training~\citep{mei2022, horvath2021}, communication and computational efficiency~\citep{wang2022}, and neural architecture search~\citep{yu2020}. Recent work such as Aequa~\citep{tastan2025aequa} further leverages slimmable networks for fair federated learning by allocating model capacity (width) to participants based on their contributions. To the best of our knowledge, our work is the first to exploit this paradigm for addressing rank heterogeneity in federated LoRA.

\section{Preliminaries}

\subsection{Federated Learning Setup}
We consider a standard federated learning (FL) setting with $K$ clients, each holding a private local dataset $\mathcal{D}_k$ of size $n_k$, with $N = \sum_k n_k$ denoting the total data size, where $k \in K$. In each communication round, a subset of clients performs local training and uploads their updates to a central server, which aggregates them into a global model and redistributes it to the clients. The goal is to minimize the global objective:
\begin{equation}
\min_W F(W) = \sum_k \frac{n_k}{N} F_k(W),
\end{equation}
where $F_k(W)$ is the local empirical loss on $\mathcal{D}_k$.

\subsection{Low-Rank Adaptation}
Fine-tuning all parameters of a large pre-trained model $W_0 \in \mathbb{R}^{m \times n}$ is often infeasible in FL due to communication and memory constraints. LoRA~\citep{hu2022lora} addresses this by freezing $W_0$ and representing the weight update as a low-rank decomposition:
\begin{equation}
W' = W_0 + \Delta W = W_0 + BA,
\end{equation}
where $B \in \mathbb{R}^{m \times r}$, $A \in \mathbb{R}^{r \times n}$, and $r \ll \min(m, n)$. Only $A$ and $B$ are trained, significantly reducing the number of trainable parameters. The rank $r$ controls the capacity of the update, trading off expressivity and resource cost.

\subsection{LoRA Aggregation in Federated Learning}
In federated LoRA, each client $k \in K$ locally trains its adapter parameters $(A_k, B_k)$ and sends them to the server for aggregation. In the homogeneous setting, a straightforward approach is to average the individual adapters:
\begin{equation}
A^{\text{global}} = \sum_k \frac{n_k}{N} A_k, \qquad
B^{\text{global}} = \sum_k \frac{n_k}{N} B_k.
\end{equation}

However, the product of the averaged LoRA adapters is not equal to the average of the corresponding low-rank updates $B_k A_k$. In particular,

\begin{equation}
\underbrace{\sum_k \frac{n_k}{N} B_k A_k}_{\Delta W^{\text{ideal}}}
\;\neq\;
\underbrace{
B^{\text{global}} A^{\text{global}}
=
\left(\sum_k \frac{n_k}{N} B_k\right)
\left(\sum_k \frac{n_k}{N} A_k\right)
}_{\Delta W^{\text{avg}}}
\end{equation}

The right-hand side introduces cross client terms $B_i A_j$ for $i \neq j$~\citep{wang2024flora, singhal-etal-2025-fedex}, highlighting the challenge of aggregating decomposed updates in federated settings.

\subsection{Aggregation in the Gradient Space}

In practice, clients may operate under different resource constraints and thus use different LoRA ranks $r_k$, leading to heterogeneous adapter dimensions across clients. This makes direct aggregation challenging and often requires additional handling such as zero padding of low-rank clients.

Additionally, to avoid cross-client interaction terms introduced by factor-wise averaging, we perform aggregation in the update space via pseudo-gradients. Each client computes gradients with respect to its local adapter parameters, $\nabla A_k$ and $\nabla B_k$, and shares these with the server. The server then performs a aggregation following the segment-wise aggregation in Section \ref{sec:segment_aggregation}
and applies these aggregated gradients to obtain the global adapter parameters $A^{\text{global}}$ and $B^{\text{global}}$.

This formulation avoids explicit multiplication of independently averaged factors and thus eliminates cross terms of the form $B_i A_j$ for $i \neq j$, ensuring that only valid client-specific update directions contribute to the global model.

\section{Method}

We address rank heterogeneity in federated LoRA by jointly designing a nested training strategy and a segment-wise aggregation rule. The key idea is to ensure that lower-rank dimensions encode meaningful and shared representations across clients, while higher-rank dimensions capture additional capacity. To achieve this, we train each client to make its adapter useful under multiple truncations, encouraging task-relevant signal to concentrate in low-rank prefix dimensions. We then aggregate adapters in a segment-wise manner, combining only the clients that contribute to each rank segment. This coupling between training and aggregation enables consistent knowledge sharing across heterogeneous ranks.

\subsection{Nested Low-Rank Structure}
We impose a prefix-structured organization over adapter dimensions. 
For a client with rank $r_k$, only the first $r_k$ dimensions are active. 
During local training, however, the adapter is optimized not only at full rank but also under multiple truncated configurations, where only prefix subsets of dimensions are used.

This ensures that lower-rank segments within a high-rank adapter are individually trained to be effective, rather than relying solely on higher dimensions. As a result, prefix dimensions encode meaningful representations that remain useful across different rank levels. As illustrated in  Figure~\ref{fig:nested_method}, a higher-rank adapter naturally contains multiple lower-rank sub-adapters within its leading dimensions.

\subsection{Local Training Strategy}
To ensure that prefix dimensions encode meaningful representations, we train each client to perform well not only at its full rank but also at the lower rank levels that appear in the federation.

Let $\mathcal{R} = \{r_1 < r_2 < \cdots < r_L\}$ denote the set of distinct rank levels in the system. For a client with rank $r_k$, we define its active training ranks as
\begin{equation}
\mathcal{R}_k = \{r \in \mathcal{R} \mid r \le r_k\}.
\end{equation}
Thus, a client is trained at its full rank as well as all lower rank levels that may participate in aggregation with it.

For each $r \in \mathcal{R}_k$, we construct a truncated adapter by activating only the first $r$ dimensions and masking the remaining dimensions. Denoting the corresponding model by $f(\cdot;\theta, r)$, where $\theta$ includes the shared frozen backbone and LoRA parameters, the local objective for client $k$ is


\begin{equation}
\mathcal{L}_k
=
\frac{1}{|\mathcal{R}_k|}
\sum_{r \in \mathcal{R}_k}
\mathcal{L}_{\mathrm{CE}}\big(f(\cdot;\theta, r), \mathcal{D}_k\big),
\end{equation}


where $\mathcal{L}_{\mathrm{CE}}$ denotes the standard cross-entropy loss.

In practice, each training step consists of one full-rank forward/backward pass and additional sub-rank passes for all $r \in \mathcal{R}_k \setminus \{r_k\}$. This training strategy ensures that each prefix segment is explicitly optimized at the same rank granularity used later during aggregation, encouraging task-relevant signal to concentrate in lower-rank dimensions. As a result, prefix segments remain informative and directly compatible with segment-wise aggregation across heterogeneous clients.


\begin{figure}[t]
    \centering
    \includegraphics[width=0.95\linewidth]{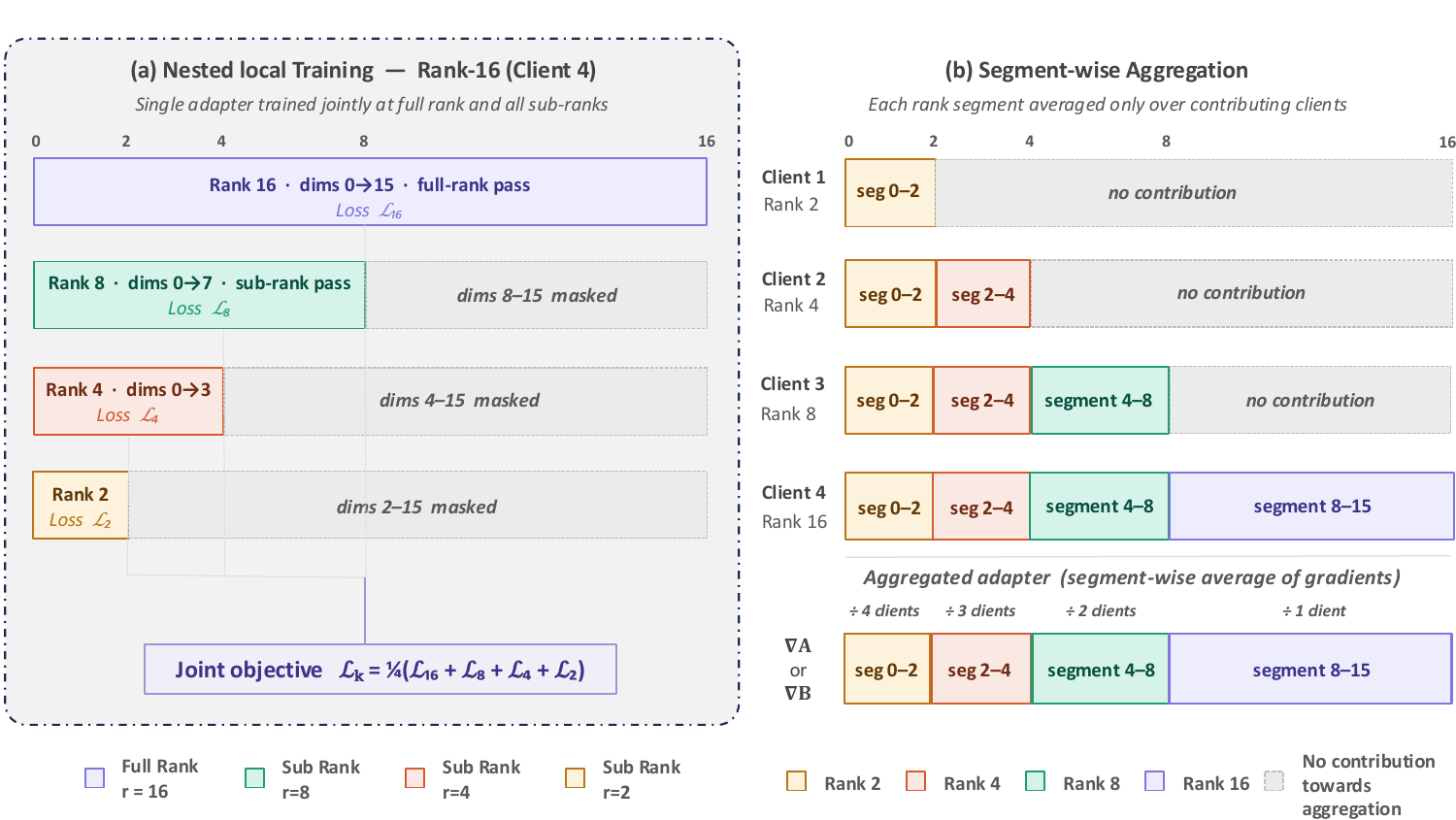}
    \caption{
Overview of the proposed nested federated LoRA method using an illustrative example with ranks {2, 4, 8, 16}. 
Left: nested local training optimizes a single adapter at full rank along with multiple prefix sub-ranks, encouraging lower-rank dimensions to capture task-relevant information through a joint objective. 
Right: segment-wise aggregation in the update space, where each client shares pseudo-gradients ($\nabla A_k, \nabla B_k$) for its active rank segments. The server performs segment-wise averaging of these gradients over contributing clients and applies the aggregated updates to obtain the global adapter parameters ($A^{\text{global}}, B^{\text{global}}$).
}
    \label{fig:nested_method}
\end{figure}

The total local objective is the average over all $|\mathcal{R}_k|$ sub-rank losses. This encourages the prefix dimensions to concentrate the most task-relevant signal, while higher dimensions encode incremental refinements, as depicted in  Figure~\ref{fig:nested_method}. The resulting adapter is thus meaningful at every prefix truncation, making it compatible with aggregation across clients of varying ranks.

\subsection{Segment-wise Aggregation}
\label{sec:segment_aggregation}

Since clients contribute only to rank segments within their capacity, standard averaging across all clients requires aligning heterogeneous adapters, typically via zero-padding lower-rank adapters to the maximum rank. This introduces zero-valued contributions in higher-rank segments for low-rank clients, diluting updates from high-rank clients. Instead, we aggregate each rank segment independently in the update space, averaging only over clients whose rank is sufficient to contribute to that segment.

Formally, let $\{r_1 < r_2 < \cdots < r_L\}$ denote the set of distinct rank levels across clients. For the $l$-th segment $[r_{l-1}, r_l)$, clients share pseudo-gradients $\nabla A_k$ and $\nabla B_k$, and the server computes:
\begin{equation}
\nabla A^{\text{global}}_{r_{l-1}:r_l} =
\frac{\displaystyle\sum_{k:\, r_k \geq r_l} n_k\, \nabla A^{k}_{r_{l-1}:r_l}}{\displaystyle\sum_{k:\, r_k \geq r_l} n_k},
\end{equation}
and analogously for $\nabla B^{\text{global}}_{r_{l-1}:r_l}$. The aggregated gradients are then applied to obtain $A^{\text{global}}$ and $B^{\text{global}}$.

This aligns with training: each segment $[r_{l-1}, r_l)$ is optimized only by clients with rank at least $r_l$, so the same clients contribute during aggregation, ensuring consistency.

As illustrated in Figure~\ref{fig:nested_method}, segment $0$–$2$ aggregates over all clients, $2$–$4$ over three, $4$–$8$ over two, and $8$–$16$ over only the highest-rank client, preventing signal dilution while preserving meaningful lower-rank sharing.


\section{Experiments}

\subsection{Experimental Setup}

\paragraph{Datasets and Setup.}
We evaluate on two standard instruction-tuning benchmarks: \textbf{Alpaca} (52K examples)~\citep{alpaca} and \textbf{Databricks-dolly-15} ~\citep{zhang2024buildingfederatedgptfederated}, as well as a classification dataset, \textbf{20 Newsgroups} ~\citep{twenty_newsgroups_113}. 

We evaluate both classification and generation performance. Specifically, we report accuracy on MMLU (Dolly, Alpaca) and 20 Newsgroups, and Rouge-L on instruction-tuning benchmarks, as shown in Table~\ref{tab:mmlu_newsgroup_results}. We additionally report perplexity to measure token-level likelihood, as shown in Table~\ref{tab:main_large_models}.

We evaluate on two base models: \textbf{TinyLlama-1.1B}~\citep{zhang2024tinyllamaopensourcesmalllanguage} and \textbf{Qwen2.5-0.5B} ~\citep{qwen2025qwen25technicalreport}. All experiments run for 15 communication rounds.

\paragraph{Implementation Details.}
For both TinyLlama-1.1B and Qwen2.5-0.5B, LoRA modules are applied to self-attention layers only following \citet{hu2022lora}. Local training uses the AdamW optimizer with a learning rate of $3 \times 10^{-4}$, batch size of 16, and gradient accumulation over 4 steps (effective batch size of 64). Clients are sampled uniformly at random in each round. Each selected client performs two local epochs per round (unless otherwise specified), and the datasets are uniformly partitioned between clients using a fixed random split. All experiments are conducted on a single NVIDIA GPU with 49GB of memory. Training each experiment takes approximately 10-17 hours, depending on the model and dataset. We did not use distributed large-scale training or specialized hardware.

\paragraph{Baselines.}
We compare against representative heterogeneous federated LoRA methods:
\textbf{ZeroPad}~\citep{zhang2024buildingfederatedgptfederated, wang2024flora}, which applies FedAvg with zero-padding to align heterogeneous ranks;
\textbf{HetLoRA}~\citep{cho-etal-2024-heterogeneous}, which uses rank self-pruning with sparsity-weighted aggregation; and
\textbf{FLoRA}~\citep{wang2024flora}, which employs stacking-based aggregation for noise-free updates.

We also report a \textbf{homogeneous reference curve}, where all clients use a fixed rank $r \in \{1,2,4,8,16,32,64,128\}$. This provides a reference for understanding the effective rank--performance trade-off under uniform capacity.

In the heterogeneous LoRA setting, we follow \citep{wang2024flora} and adopt their rank configuration for the 10-client case, assigning local ranks [64, 32, 16, 16, 8, 8, 4, 4, 4, 4]. For the 5-client setting, we use a similar rank pattern, i.e., [16, 8, 4, 2, 1], to simulate heterogeneous computational resources. Nested training evaluates all valid prefix sub-ranks for each batch.

\begin{table*}[t]
\centering
\small
\setlength{\tabcolsep}{6pt}
\renewcommand{\arraystretch}{1.12}
\caption{
Evaluation of heterogeneous federated LoRA methods across base models. 
Accuracy is reported for MMLU (Dolly, Alpaca) and 20 Newsgroups classification, 
while Rouge-L is reported for instruction-tuning benchmarks (Dolly, Alpaca). 
Higher is better.
}
\label{tab:mmlu_newsgroup_results}
\begin{tabular}{llccc cc}
\toprule
\textbf{Foundation Model} & \textbf{Method} 
& \multicolumn{3}{c}{\textbf{Accuracy}} 
& \multicolumn{2}{c}{\textbf{ROUGE-L}} \\
\cmidrule(lr){3-5} \cmidrule(lr){6-7}
& 
& \multicolumn{2}{c}{\textbf{MMLU}} 
& \textbf{} 
& \multicolumn{2}{c}{\textbf{}} \\
\cmidrule(lr){3-4} 
& 
& \textbf{Dolly} & \textbf{Alpaca} 
& \textbf{20 Newsgroups} 
& \textbf{Dolly} & \textbf{Alpaca} \\
\midrule

\multirow{6}{*}{\textbf{TinyLlama-1.1B}}
& ZeroPad              & 29.1 & 26.72 & 14.68 & 27.34 & 32.1 \\
& HetLoRA              & 29.9 & 29.09 & 17.01 & 28.13 & \underline{33.27} \\
& FLoRA                & 28.42 & 29.19 & \textbf{34.08} & \underline{28.72} & 33.21 \\
& Nested Aggregation Only & \underline{29.63} & 28.63 & 29.48 & 28.16 & 28.76 \\
& Nested Training Only & 28.73 & \underline{30.21} & 31.68 & 26.65 & 33.2 \\
& \method                 &\textbf{ 33.86} & \textbf{30.76} & \underline{32.47} & \textbf{29.34} & \textbf{33.79} \\
\midrule

\multirow{6}{*}{\textbf{Qwen2.5-0.5B}}
& ZeroPad              & 38.36 & 33.1 & 16.95 & 27.07 & 30.49 \\
& HetLoRA              & 38.88 & 34.59 & 30.90 & 27.23 & 32.47 \\
& FLoRA                & 38.84 & 34.51 & 32.94 & 28.33 & 34.22 \\
& Nested Aggregation Only & 38.32 & 32.71 & \textbf{39.93} & \underline{28.42} & 30.6 \\
& Nested Training Only & \underline{39.33} & \underline{34.66} & 28.22 & 26.53 & \underline{34.43} \\
& \method                 & \textbf{39.85} & \textbf{35.76} & \underline{39.52} & \textbf{28.79} & \textbf{35.03} \\

\bottomrule
\end{tabular}
\end{table*}

\begin{table*}[t]
\centering
\small
\setlength{\tabcolsep}{3.5pt}
\renewcommand{\arraystretch}{1.08}

\definecolor{sectionpink}{RGB}{239,205,205}
\definecolor{selfblue}{RGB}{214,228,238}

\caption{
Perplexity comparison of heterogeneous federated LoRA methods on instruction-tuning benchmarks using TinyLlama-1.1B and Qwen2.5-0.5B. Lower is better.}
\label{tab:main_large_models}

\begin{tabular}{lcccccc}
\toprule
\textbf{Method} & \textbf{ZeroPad} & \textbf{HetLoRA} & \textbf{FLoRA} & \textbf{Nested Aggregation Only} & \textbf{Nested Training Only} & \textbf{\method}  \\
\midrule

\rowcolor{sectionpink}
\multicolumn{7}{c}{\textbf{TinyLlama 1.1B}} \\
Dolly      & 4.5902 & 4.5589 & \underline{4.4407} & 4.9115 & 4.5492 & \textbf{4.4030}  \\
Alpaca         & 2.6822 & 2.6822 &  \textbf{2.6230} & 3.1084 & 2.6699 & \underline{2.6379}  \\
NewsGroup & 7.1240 & 6.9611 & \underline{6.5561} & 6.6924 & 6.6105 & \textbf{6.4899} \\

\midrule

\rowcolor{sectionpink}
\multicolumn{7}{c}{\textbf{Qwen-2.5-0.5B}} \\
Dolly       & 7.0482 & 7.0276 & \textbf{6.8025} & 7.6927 & 6.9159 & \underline{6.8156} \\
Alpaca          & 2.8185 & 2.7983 & 2.6774 & \underline{2.6695} & 3.4447 & \textbf{2.6378} \\

NewsGroup  & 11.6069 & 11.5412 & 11.4055 & 11.9582 & \underline{11.3023} & \textbf{11.2229} \\
\bottomrule
\end{tabular}
\end{table*}

\begin{figure}[t]
\centering
\begin{tikzpicture}
\begin{groupplot}[
    group style={group size=2 by 1, horizontal sep=0.8cm,
    x descriptions at=edge bottom,
    y descriptions at=edge left},
    width=0.48\linewidth,
    height=0.35\linewidth,
    xmode=log,
    log basis x=2,
    grid=both,
    grid style={gray!20},
    xlabel={rank},
    ylabel={perplexity},
    tick label style={font=\scriptsize},
    label style={font=\scriptsize},
    title style={font=\small},
]

\nextgroupplot[
    title={Alpaca},
    xmin=1, xmax=280,
    ymin=2.62, ymax=2.93,
    xtick={1,2,4,8,16,32,64,128,220},
    xticklabels={1,2,4,8,16,32,64,128,160},
    after end axis/.code={
        \draw[white, line width=4pt]
            ({axis cs:145,2.62}) -- ({axis cs:190,2.62});
        \draw[black, thick]
            ({axis cs:148,2.619}) -- ({axis cs:155,2.621})
            ({axis cs:160,2.619}) -- ({axis cs:167,2.621});
    },
]

\addplot+[blue, mark=*, thick, dashed] coordinates {
    (1,2.9111) (2,2.8293) (4,2.7694) (8,2.7286)
    (16,2.6941) (32,2.6661) (64,2.6442) (128,2.6257)
};
\addplot+[red, mark=*, thick] coordinates {
    (4,2.8650) (8,2.8500) (16,2.8350) (32,2.8260) (64,2.8185)
};
\addplot+[brown, mark=*, thick] coordinates {
    (4,2.8450) (8,2.8320) (16,2.8180) (32,2.8070) (64,2.7983)
};
\addplot+[blue!70!black, mark=*, thick] coordinates {
    (4,2.6543) (8,2.6498) (16,2.6459) (32,2.6439) (64,2.6695)
};
\addplot[red!80!black, mark=*, thick] coordinates {
    (4,2.6440) (8,2.6405) (16,2.6392) (32,2.6384) (64,2.6378)
};
\addplot[orange, mark=*, thick] coordinates {
    (4,2.6731) (8,2.6720) (16,2.6712) (32,2.6705) (64,2.6699)
};
\addplot[green!60!black, only marks, mark=*, mark size=2pt] coordinates {(220,2.6774)};

\foreach \y in {2.8185,2.7983,2.6774,2.6695,2.6378} {
    \addplot[blue!25, dashed, forget plot] coordinates {(1,\y) (280,\y)};
}

\nextgroupplot[
    title={Dolly},
    xmin=1, xmax=128,
    ymin=4.34, ymax=4.65,
    xtick={1,2,4,8,16,32,64,128},
    xticklabels={1,2,4,8,16,32,64,128},
]

\addplot+[blue, mark=*, thick, dashed] coordinates {
    (1,4.639) (2,4.5805) (4,4.5052) (8,4.4649)
    (16,4.4095) (32,4.371) (64,4.3473) (128,4.3473)
};
\addplot+[red, mark=*, thick] coordinates {
    (1,4.6400) (2,4.6320) (4,4.6280) (8,4.6240) (16,4.6201)
};
\addplot+[brown, mark=*, thick] coordinates {
    (1,4.6131) (2,4.5850) (4,4.5650) (8,4.5580) (16,4.5595)
};
\addplot[green!60!black, only marks, mark=*, mark size=2pt] coordinates {(31,4.3943)};
\addplot+[blue!70!black, mark=*, thick] coordinates {
    (1,4.4559) (2,4.4498) (4,4.4480) (8,4.4402) (16,4.5185)
};
\addplot[red!80!black, mark=*, thick] coordinates {
    (1,4.4155) (2,4.4052) (4,4.3975) (8,4.3948) (16,4.3964)
};
\addplot[orange, mark=*, thick] coordinates {
    (1,4.4496) (2,4.4399) (4,4.4332) (8,4.4319) (16,4.4321)
};

\foreach \y in {4.6201,4.5385,4.4243,4.5185,4.3922} {
    \addplot[blue!25, dashed, forget plot] coordinates {(1,\y) (128,\y)};
}

\end{groupplot}

\node at ($(group c1r1.south)!0.5!(group c2r1.south)$) [yshift=-1.35cm] {
\begin{tikzpicture}
\begin{axis}[
    hide axis,
    xmin=0, xmax=1,
    ymin=0, ymax=1,
    legend columns=3,
    legend style={
        draw=none,
        font=\scriptsize,
        /tikz/every even column/.append style={column sep=0.35cm}
    }
]
\addlegendimage{blue, mark=*, thick, dashed}
\addlegendentry{Homogeneous}
\addlegendimage{red, mark=*, thick}
\addlegendentry{ZeroPad}
\addlegendimage{brown, mark=*, thick}
\addlegendentry{HetLoRA}
\addlegendimage{green!60!black, only marks, mark=*}
\addlegendentry{FLoRA}
\addlegendimage{blue!70!black, mark=*, thick}
\addlegendentry{Nested Aggregation}
\addlegendimage{orange, mark=*, thick}
\addlegendentry{Nested Training}
\addlegendimage{red!80!black, mark=*, thick}
\addlegendentry{\method}
\end{axis}
\end{tikzpicture}
};

\end{tikzpicture}
\caption{
Comparison of homogeneous and heterogeneous methods on Alpaca and Dolly.
For heterogeneous methods, each curve shows performance under prefix rank 
truncation: the rightmost point corresponds to the full trained rank, and 
points to the left reflect evaluation at progressively smaller prefix sub-ranks 
of the same trained adapter. FLoRA is shown as a single point at its effective 
rank (160 for 10 client setting, 31 for 5 client setting), as it merges adapters into the backbone 
after each round and does not maintain a persistent adapter amenable to 
truncation}
\label{fig:main_results}
\end{figure}
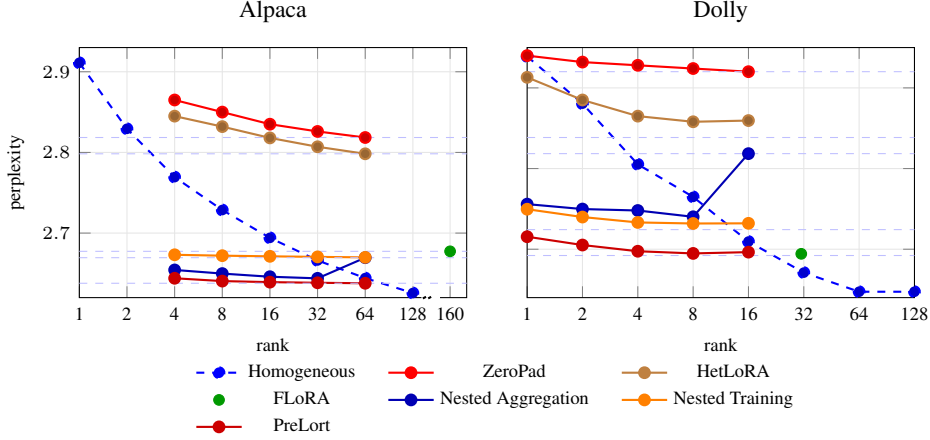

\subsection{Main Results}

Table~\ref{tab:mmlu_newsgroup_results} presents the main comparison across heterogeneous federated LoRA methods in terms of accuracy and ROUGE-L.

Across both TinyLlama-1.1B and Qwen2.5-0.5B, our method (\method) consistently achieves the best performance on MMLU (Dolly, Alpaca) and ROUGE-L metrics. In particular, it outperforms all baselines on both MMLU tasks across models, and achieves the highest ROUGE-L scores on Dolly and Alpaca. On 20 Newsgroups, \method remains competitive, achieving the second-best performance while outperforming most baselines. These results indicate that our approach improves instruction-following and generalization performance without sacrificing robustness across tasks.

Table~\ref{tab:main_large_models} reports perplexity on instruction-tuning datasets. Our method achieves the lowest perplexity on Dolly and NewsGroup for both base models, and remains competitive on a few settings, where FLoRA performs strongly. Compared to FLoRA, which benefits from stacking-based aggregation that increases effective rank, \method achieves comparable or better perplexity while maintaining a fixed parameter budget, avoiding growth in adapter size as the number of clients increases~\citep{wang2024flora}.


Overall, the results demonstrate that (i) segment-wise aggregation effectively utilizes heterogeneous rank capacity without dilution, and (ii) the nested training objective is critical for aligning representations across ranks. The ablation (Nested Aggregation Only and Nested Training Only) confirms that aggregation alone is insufficient, and that performance gains arise from the combination of both components.


\subsection{Analysis}

\subsubsection{Per-Dimension Importance}

\begin{figure}[t]
\centering

\begin{tikzpicture}
\begin{groupplot}[
    group style={
        group size=2 by 2,
        horizontal sep=1.35cm,
        vertical sep=1.75cm
    },
    width=0.44\linewidth,
    height=0.37\linewidth,
    grid=both,
    grid style={gray!20},
    xlabel={dim. index},
    ylabel={importance},
    tick label style={font=\scriptsize},
    label style={font=\scriptsize},
    title style={font=\small},
    every mark/.append style={scale=0.45},
    scaled y ticks=false,
    yticklabel style={
        /pgf/number format/fixed,
        /pgf/number format/precision=2
    },
]

\nextgroupplot[
    title={$r=16$},
    xmin=0, xmax=10,
    ymin=0.025, ymax=0.125,
    xtick={0,4,8,12,15},
    ytick={0.04,0.06,0.08,0.10,0.12},
]

\addplot[name path=b1r16upper, draw=none, forget plot] coordinates {
(0,0.0674) (1,0.0532) (2,0.0496) (3,0.0457)
(4,0.0537) (5,0.0510) (6,0.0558) (7,0.0554)
(8,0.0883) (9,0.0892) (10,0.0862) (11,0.0878)
(12,0.0887) (13,0.0870) (14,0.0913) (15,0.0901)
};
\addplot[name path=b1r16lower, draw=none, forget plot] coordinates {
(0,0.0449) (1,0.0365) (2,0.0359) (3,0.0355)
(4,0.0420) (5,0.0411) (6,0.0431) (7,0.0417)
(8,0.0654) (9,0.0670) (10,0.0677) (11,0.0698)
(12,0.0661) (13,0.0682) (14,0.0671) (15,0.0679)
};
\addplot[blue, opacity=0.14, forget plot] fill between[of=b1r16upper and b1r16lower];

\addplot[name path=m5r16upper, draw=none, forget plot] coordinates {
(0,0.1187) (1,0.0773) (2,0.0694) (3,0.0689)
(4,0.0677) (5,0.0663) (6,0.0656) (7,0.0652)
(8,0.0626) (9,0.0617) (10,0.0611) (11,0.0605)
(12,0.0590) (13,0.0566) (14,0.0552) (15,0.0522)
};
\addplot[name path=m5r16lower, draw=none, forget plot] coordinates {
(0,0.0763) (1,0.0657) (2,0.0632) (3,0.0619)
(4,0.0610) (5,0.0605) (6,0.0614) (7,0.0601)
(8,0.0573) (9,0.0566) (10,0.0558) (11,0.0545)
(12,0.0523) (13,0.0507) (14,0.0489) (15,0.0458)
};
\addplot[red, opacity=0.14, forget plot] fill between[of=m5r16upper and m5r16lower];

\addplot[gray, dotted, thick] coordinates {(0,0.0625) (15,0.0625)};
\addplot[blue, thick, mark=*, mark size=1.0pt] coordinates {
(0,0.0561) (1,0.0449) (2,0.0427) (3,0.0406)
(4,0.0479) (5,0.0460) (6,0.0495) (7,0.0485)
(8,0.0768) (9,0.0781) (10,0.0769) (11,0.0788)
(12,0.0774) (13,0.0776) (14,0.0792) (15,0.0790)
};
\addplot[red, thick, mark=*, mark size=1.0pt] coordinates {
(0,0.0975) (1,0.0715) (2,0.0663) (3,0.0654)
(4,0.0643) (5,0.0634) (6,0.0635) (7,0.0626)
(8,0.0599) (9,0.0592) (10,0.0585) (11,0.0575)
(12,0.0556) (13,0.0536) (14,0.0521) (15,0.0490)
};

\nextgroupplot[
    title={$r=8$},
    xmin=0, xmax=7,
    ymin=0.08, ymax=0.22,
    xtick={0,2,4,6,7},
    ytick={0.10,0.15,0.20},
    ylabel={},
]

\addplot[name path=b1r8upper, draw=none, forget plot] coordinates {
(0,0.1627) (1,0.1333) (2,0.1319) (3,0.1191)
(4,0.1429) (5,0.1395) (6,0.1508) (7,0.1477)
};
\addplot[name path=b1r8lower, draw=none, forget plot] coordinates {
(0,0.1206) (1,0.0989) (2,0.0956) (3,0.0968)
(4,0.1164) (5,0.1113) (6,0.1179) (7,0.1145)
};
\addplot[blue, opacity=0.14, forget plot] fill between[of=b1r8upper and b1r8lower];

\addplot[name path=m5r8upper, draw=none, forget plot] coordinates {
(0,0.2161) (1,0.1446) (2,0.1298) (3,0.1251)
(4,0.1225) (5,0.1193) (6,0.1162) (7,0.1128)
};
\addplot[name path=m5r8lower, draw=none, forget plot] coordinates {
(0,0.1530) (1,0.1258) (2,0.1161) (3,0.1129)
(4,0.1071) (5,0.1023) (6,0.0996) (7,0.0969)
};
\addplot[red, opacity=0.14, forget plot] fill between[of=m5r8upper and m5r8lower];

\addplot[gray, dotted, thick] coordinates {(0,0.125) (7,0.125)};
\addplot[blue, thick, mark=*, mark size=1.0pt] coordinates {
(0,0.1417) (1,0.1161) (2,0.1137) (3,0.1080)
(4,0.1297) (5,0.1254) (6,0.1344) (7,0.1311)
};
\addplot[red, thick, mark=*, mark size=1.0pt] coordinates {
(0,0.1845) (1,0.1352) (2,0.1230) (3,0.1190)
(4,0.1148) (5,0.1108) (6,0.1079) (7,0.1048)
};

\nextgroupplot[
    title={$r=4$},
    xmin=0, xmax=3,
    ymin=0.17, ymax=0.39,
    xtick={0,1,2,3},
    ytick={0.20,0.25,0.30,0.35},
]

\addplot[name path=b1r4upper, draw=none, forget plot] coordinates {
(0,0.3238) (1,0.2727) (2,0.2734) (3,0.2549)
};
\addplot[name path=b1r4lower, draw=none, forget plot] coordinates {
(0,0.2577) (1,0.2137) (2,0.2030) (3,0.2009)
};
\addplot[blue, opacity=0.14, forget plot] fill between[of=b1r4upper and b1r4lower];

\addplot[name path=m5r4upper, draw=none, forget plot] coordinates {
(0,0.3755) (1,0.2602) (2,0.2322) (3,0.2232)
};
\addplot[name path=m5r4lower, draw=none, forget plot] coordinates {
(0,0.2954) (1,0.2259) (2,0.2001) (3,0.1875)
};
\addplot[red, opacity=0.14, forget plot] fill between[of=m5r4upper and m5r4lower];

\addplot[gray, dotted, thick] coordinates {(0,0.25) (3,0.25)};
\addplot[blue, thick, mark=*, mark size=1.0pt] coordinates {
(0,0.2907) (1,0.2432) (2,0.2382) (3,0.2279)
};
\addplot[red, thick, mark=*, mark size=1.0pt] coordinates {
(0,0.3354) (1,0.2430) (2,0.2162) (3,0.2053)
};

\nextgroupplot[
    title={$r=2$},
    xmin=0, xmax=1,
    ymin=0.34, ymax=0.64,
    xtick={0,1},
    ytick={0.40,0.50,0.60},
    ylabel={},
]

\addplot[name path=b1r2upper, draw=none, forget plot] coordinates {
(0,0.5870) (1,0.4988)
};
\addplot[name path=b1r2lower, draw=none, forget plot] coordinates {
(0,0.5012) (1,0.4130)
};
\addplot[blue, opacity=0.14, forget plot] fill between[of=b1r2upper and b1r2lower];

\addplot[name path=m5r2upper, draw=none, forget plot] coordinates {
(0,0.6243) (1,0.4543)
};
\addplot[name path=m5r2lower, draw=none, forget plot] coordinates {
(0,0.5457) (1,0.3757)
};
\addplot[red, opacity=0.14, forget plot] fill between[of=m5r2upper and m5r2lower];

\addplot[gray, dotted, thick] coordinates {(0,0.5) (1,0.5)};
\addplot[blue, thick, mark=*, mark size=1.0pt] coordinates {
(0,0.5441) (1,0.4559)
};
\addplot[red, thick, mark=*, mark size=1.0pt] coordinates {
(0,0.5850) (1,0.4150)
};

\end{groupplot}
\end{tikzpicture}

\vspace{0.55em}

\centering
\begin{tikzpicture}
\begin{axis}[
    hide axis,
    xmin=0, xmax=1,
    ymin=0, ymax=1,
    legend columns=3,
    legend style={
        draw=none,
        font=\scriptsize,
        /tikz/every even column/.append style={column sep=0.8cm}
    }
]
\addlegendimage{blue, thick, mark=*, mark size=1.0pt}
\addlegendentry{Nested aggregation only}

\addlegendimage{red, thick, mark=*, mark size=1.0pt}
\addlegendentry{\method}

\addlegendimage{gray, dotted, thick}
\addlegendentry{Uniform}
\end{axis}
\end{tikzpicture}

\vspace{-0.6em}

\caption{
Per-dimension importance across rank budgets.
Nested aggregation only and our method are shown for ranks $16$, $8$, $4$, and $2$.
Shaded regions denote one standard deviation, and the dotted line denotes the uniform importance baseline $1/r$.
}
\label{fig:perdim}
\end{figure}
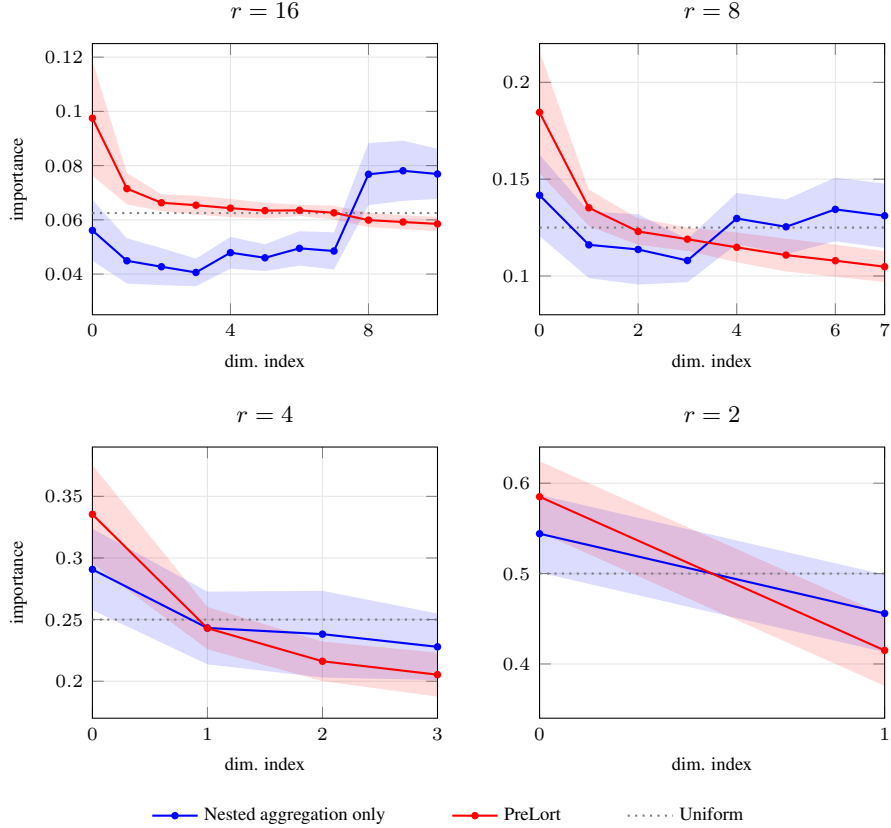

To better understand how nested training affects the internal structure of the adapter, we analyze the per-dimension importance of each rank index $i$. We define importance as the product of the Frobenius norm of the $i$-th row of $A$ and the $i$-th column of $B$, normalized across all dimensions. This provides a proxy for the contribution of each rank component to the overall update $\Delta W = BA$.

As shown in Figure~\ref{fig:perdim}, standard training yields an approximately uniform distribution of importance across all 16 dimensions, closely matching the $1/16$ baseline. In contrast, our approach produces a strongly structured distribution, where importance is concentrated in the leading dimensions. This behavior suggests that the adapter learns an implicit prioritization over rank components. Such a structure is particularly beneficial for nested aggregation. Low-rank clients contribute to these high-impact dimensions, while higher-rank clients provide incremental refinements, leading to more stable and effective updates.


\subsubsection{Rank Robustness}

Figure~\ref{fig:main_results} further illustrates the rank–performance trade-off. It shows the performance of each method across different effective rank levels. The homogeneous reference improves steadily as rank increases, as expected. In contrast, our method remains stable across the lower-rank regime and achieves consistently lower perplexity than the heterogeneous baselines on both Alpaca and Dolly.

\paragraph{Evaluation Protocol.}
For \textbf{FLoRA}, adapter updates are merged into the frozen backbone after each communication round, and fresh adapters are initialized in subsequent rounds. As a result, there is no persistent aggregated adapter at the end of training, and we report a single perplexity value corresponding to the final merged model. The effective rank of FLoRA corresponds to the stacked adapter rank accumulated during training, which equals the sum of client ranks. In our setup, this results in an effective rank of 160 for the Alpaca (10-client) setting and 31 for the Dolly (5-client) setting.

In contrast, methods such as ZeroPad, HetLoRA, Nested Aggregation, and \method\ maintain an explicit aggregated adapter after each round of aggregation on the server. This allows us to evaluate rank robustness by truncating the final aggregated adapter to prefix ranks at evaluation time. The reported curves for these methods therefore reflect the performance of the same trained adapter under different truncation levels. 

The \textit{Nested Aggregation only} baseline exhibits unstable behavior, particularly at higher ranks, indicating that aggregation alone is insufficient. Without nested training, importance is not aligned with the aggregation structure: higher-rank dimensions can dominate locally but are supported by only a subset of clients during aggregation. As illustrated in Figure~\ref{fig:perdim}, these dimensions exhibit high importance despite limited support, leading to overfitting and instability.

In contrast, our nested training enforces a hierarchical ordering of importance, concentrating task-relevant signal in shared prefix dimensions. This acts as an implicit regularizer and yields stable performance across rank levels.

\subsection{Limitations.}
The effectiveness of the proposed nested training strategy depends on sufficient local optimization steps; with limited local epochs, the model may not fully learn well-aligned low-rank prefix representations. But this also improves communication efficiency by converging faster. Nested training introduces additional computational overhead due to multiple forward/backward passes per batch, increasing training time by approximately 40\% in our setup, although per-step compute efficiency remains unchanged. We observe this overhead to be the same in our experimental settings, though its behavior at larger scales is not explicitly evaluated.

\vspace{-0.2cm}

\section{Conclusion}

\vspace{-0.2cm}

We introduced \method, a federated LoRA fine-tuning method designed for rank-heterogeneous clients. The core idea is to make heterogeneous ranks compatible not just at aggregation time, but during training itself. To this end, nested training optimizes each adapter under multiple prefix truncations, forcing task-relevant information to concentrate in shared low-rank dimensions. Segment-wise aggregation then averages each rank segment only over clients that actually contribute to it, avoiding zero-padding dilution while preserving useful higher-rank capacity.

Across Alpaca, Dolly, and 20 Newsgroups with TinyLlama-1.1B and Qwen2.5-0.5B, \method consistently improves accuracy and ROUGE-L over heterogeneous LoRA baselines, while achieving lower or comparable perplexity. The ablation and per-dimension analysis further show that aggregation alone is not enough: the gains come from coupling segment-wise aggregation with nested training, which produces aligned, robust prefix representations across ranks.

Overall, our results show that rank heterogeneity in federated LoRA is best handled by jointly structuring how adapters are trained and how their updates are aggregated.
\bibliographystyle{plainnat}
\bibliography{references}

@misc{zhang2024buildingfederatedgptfederated,
      title={Towards Building the Federated GPT: Federated Instruction Tuning}, 
      author={Jianyi Zhang and Saeed Vahidian and Martin Kuo and Chunyuan Li and Ruiyi Zhang and Tong Yu and Yufan Zhou and Guoyin Wang and Yiran Chen},
      year={2024},
      eprint={2305.05644},
      archivePrefix={arXiv},
      primaryClass={cs.CL},
      url={https://arxiv.org/abs/2305.05644}, 
}

@inproceedings{
hu2022lora,
title={Lo{RA}: Low-Rank Adaptation of Large Language Models},
author={Edward J Hu and yelong shen and Phillip Wallis and Zeyuan Allen-Zhu and Yuanzhi Li and Shean Wang and Lu Wang and Weizhu Chen},
booktitle={International Conference on Learning Representations},
year={2022},
url={https://openreview.net/forum?id=nZeVKeeFYf9}
}

@inproceedings{
wang2024flora,
title={{FL}o{RA}: Federated Fine-Tuning Large Language Models with Heterogeneous Low-Rank Adaptations},
author={Ziyao Wang and Zheyu Shen and Yexiao He and Guoheng Sun and Hongyi Wang and Lingjuan Lyu and Ang Li},
booktitle={The Thirty-eighth Annual Conference on Neural Information Processing Systems},
year={2024},
url={https://openreview.net/forum?id=TcCorXxNJQ}
}

@inproceedings{singhal-etal-2025-fedex,
    title = "{F}ed{E}x-{L}o{RA}: Exact Aggregation for Federated and Efficient Fine-Tuning of Large Language Models",
    author = "Singhal, Raghav  and
      Ponkshe, Kaustubh  and
      Vepakomma, Praneeth",
    editor = "Che, Wanxiang  and
      Nabende, Joyce  and
      Shutova, Ekaterina  and
      Pilehvar, Mohammad Taher",
    booktitle = "Proceedings of the 63rd Annual Meeting of the Association for Computational Linguistics (Volume 1: Long Papers)",
    month = jul,
    year = "2025",
    address = "Vienna, Austria",
    publisher = "Association for Computational Linguistics",
    url = "https://aclanthology.org/2025.acl-long.67/",
    doi = "10.18653/v1/2025.acl-long.67",
    pages = "1316--1336",
    ISBN = "979-8-89176-251-0"
}

@inproceedings{cho-etal-2024-heterogeneous,
    title = "Heterogeneous {L}o{RA} for Federated Fine-tuning of On-Device Foundation Models",
    author = "Cho, Yae Jee  and
      Liu, Luyang  and
      Xu, Zheng  and
      Fahrezi, Aldi  and
      Joshi, Gauri",
    editor = "Al-Onaizan, Yaser  and
      Bansal, Mohit  and
      Chen, Yun-Nung",
    booktitle = "Proceedings of the 2024 Conference on Empirical Methods in Natural Language Processing",
    month = nov,
    year = "2024",
    address = "Miami, Florida, USA",
    publisher = "Association for Computational Linguistics",
    url = "https://aclanthology.org/2024.emnlp-main.717/",
    doi = "10.18653/v1/2024.emnlp-main.717",
    pages = "12903--12913"
}

@misc{zhang2024tinyllamaopensourcesmalllanguage,
      title={TinyLlama: An Open-Source Small Language Model}, 
      author={Peiyuan Zhang and Guangtao Zeng and Tianduo Wang and Wei Lu},
      year={2024},
      eprint={2401.02385},
      archivePrefix={arXiv},
      primaryClass={cs.CL},
      url={https://arxiv.org/abs/2401.02385}, 
}

@inproceedings{
kusupati2022matryoshka,
title={Matryoshka Representation Learning},
author={Aditya Kusupati and Gantavya Bhatt and Aniket Rege and Matthew Wallingford and Aditya Sinha and Vivek Ramanujan and William Howard-Snyder and Kaifeng Chen and Sham M. Kakade and Prateek Jain and Ali Farhadi},
booktitle={Advances in Neural Information Processing Systems},
editor={Alice H. Oh and Alekh Agarwal and Danielle Belgrave and Kyunghyun Cho},
year={2022},
url={https://openreview.net/forum?id=9njZa1fm35}
}

@InProceedings{pmlr-v54-mcmahan17a,
  title = 	 {{Communication-Efficient Learning of Deep Networks from Decentralized Data}},
  author = 	 {McMahan, Brendan and Moore, Eider and Ramage, Daniel and Hampson, Seth and Arcas, Blaise Aguera y},
  booktitle = 	 {Proceedings of the 20th International Conference on Artificial Intelligence and Statistics},
  pages = 	 {1273--1282},
  year = 	 {2017},
  editor = 	 {Singh, Aarti and Zhu, Jerry},
  volume = 	 {54},
  series = 	 {Proceedings of Machine Learning Research},
  month = 	 {20--22 Apr},
  publisher =    {PMLR},
  url = 	 {https://proceedings.mlr.press/v54/mcmahan17a.html}
}

@inproceedings{houlsby2019parameter,
  title = {Parameter-Efficient Transfer Learning for {NLP}},
  author = {Houlsby, Neil and Giurgiu, Andrei and Jastrzebski, Stanislaw and Morrone, Bruna and De Laroussilhe, Quentin and Gesmundo, Andrea and Attariyan, Mona and Gelly, Sylvain},
  booktitle = {Proceedings of the 36th International Conference on Machine Learning},
  year = {2019},
}

@inproceedings{pfeiffer-etal-2021-adapterfusion,
    title = "{A}dapter{F}usion: Non-Destructive Task Composition for Transfer Learning",
    author = {Pfeiffer, Jonas  and
      Kamath, Aishwarya  and
      R{\"u}ckl{\'e}, Andreas  and
      Cho, Kyunghyun  and
      Gurevych, Iryna},
    editor = "Merlo, Paola  and
      Tiedemann, Jorg  and
      Tsarfaty, Reut",
    booktitle = "Proceedings of the 16th Conference of the European Chapter of the Association for Computational Linguistics: Main Volume",
    month = apr,
    year = "2021",
    address = "Online",
    publisher = "Association for Computational Linguistics",
    url = "https://aclanthology.org/2021.eacl-main.39/",
    doi = "10.18653/v1/2021.eacl-main.39",
    pages = "487--503"
}

@inproceedings{lester-etal-2021-power,
    title = "The Power of Scale for Parameter-Efficient Prompt Tuning",
    author = "Lester, Brian  and
      Al-Rfou, Rami  and
      Constant, Noah",
    editor = "Moens, Marie-Francine  and
      Huang, Xuanjing  and
      Specia, Lucia  and
      Yih, Scott Wen-tau",
    booktitle = "Proceedings of the 2021 Conference on Empirical Methods in Natural Language Processing",
    month = nov,
    year = "2021",
    address = "Online and Punta Cana, Dominican Republic",
    publisher = "Association for Computational Linguistics",
    url = "https://aclanthology.org/2021.emnlp-main.243/",
    doi = "10.18653/v1/2021.emnlp-main.243",
    pages = "3045--3059"
}

@inproceedings{li-liang-2021-prefix,
    title = "Prefix-Tuning: Optimizing Continuous Prompts for Generation",
    author = "Li, Xiang Lisa  and
      Liang, Percy",
    editor = "Zong, Chengqing  and
      Xia, Fei  and
      Li, Wenjie  and
      Navigli, Roberto",
    booktitle = "Proceedings of the 59th Annual Meeting of the Association for Computational Linguistics and the 11th International Joint Conference on Natural Language Processing (Volume 1: Long Papers)",
    month = aug,
    year = "2021",
    address = "Online",
    publisher = "Association for Computational Linguistics",
    url = "https://aclanthology.org/2021.acl-long.353/",
    doi = "10.18653/v1/2021.acl-long.353",
    pages = "4582--4597"
}

@inproceedings{ben-zaken-etal-2022-bitfit,
    title = "{B}it{F}it: Simple Parameter-efficient Fine-tuning for Transformer-based Masked Language-models",
    author = "Ben Zaken, Elad  and
      Goldberg, Yoav  and
      Ravfogel, Shauli",
    editor = "Muresan, Smaranda  and
      Nakov, Preslav  and
      Villavicencio, Aline",
    booktitle = "Proceedings of the 60th Annual Meeting of the Association for Computational Linguistics (Volume 2: Short Papers)",
    month = may,
    year = "2022",
    address = "Dublin, Ireland",
    publisher = "Association for Computational Linguistics",
    url = "https://aclanthology.org/2022.acl-short.1/",
    doi = "10.18653/v1/2022.acl-short.1",
    pages = "1--9"
}

@inproceedings{
zhang2023adaptive,
title={Adaptive Budget Allocation for Parameter-Efficient Fine-Tuning },
author={Qingru Zhang and Minshuo Chen and Alexander Bukharin and Pengcheng He and Yu Cheng and Weizhu Chen and Tuo Zhao},
booktitle={The Eleventh International Conference on Learning Representations },
year={2023},
url={https://openreview.net/forum?id=lq62uWRJjiY}
}

@inproceedings{
liu2024dora,
title={Do{RA}: Weight-Decomposed Low-Rank Adaptation},
author={Liu, Shih-yang and Chien-Yi Wang and Hongxu Yin and Pavlo Molchanov and Yu-Chiang Frank Wang and Kwang-Ting Cheng and Min-Hung Chen},
booktitle={Forty-first International Conference on Machine Learning},
year={2024},
url={https://openreview.net/forum?id=3d5CIRG1n2}
}

@article{kairouz,
author = {Kairouz, Peter and McMahan, H. Brendan and Avent, Brendan and Bellet, Aur\'{e}lien and Bennis, Mehdi and Nitin Bhagoji, Arjun and Bonawitz, Kallista and Charles, Zachary and Cormode, Graham and Cummings, Rachel and D'Oliveira, Rafael G. L. and Eichner, Hubert and El Rouayheb, Salim and Evans, David and Gardner, Josh and Garrett, Zachary and Gasc\'{o}n, Adri\`{a} and Ghazi, Badih and Gibbons, Phillip B. and Gruteser, Marco and Harchaoui, Zaid and He, Chaoyang and He, Lie and Huo, Zhouyuan and Hutchinson, Ben and Hsu, Justin and Jaggi, Martin and Javidi, Tara and Joshi, Gauri and Khodak, Mikhail and Konecn\'{y}, Jakub and Korolova, Aleksandra and Koushanfar, Farinaz and Koyejo, Sanmi and Lepoint, Tancr\`{e}de and Liu, Yang and Mittal, Prateek and Mohri, Mehryar and Nock, Richard and \"{O}zg\"{u}r, Ayfer and Pagh, Rasmus and Qi, Hang and Ramage, Daniel and Raskar, Ramesh and Raykova, Mariana and Song, Dawn and Song, Weikang and Stich, Sebastian U. and Sun, Ziteng and Suresh, Ananda Theertha and Tram\`{e}r, Florian and Vepakomma, Praneeth and Wang, Jianyu and Xiong, Li and Xu, Zheng and Yang, Qiang and Yu, Felix X. and Yu, Han and Zhao, Sen},
title = {Advances and Open Problems in Federated Learning},
year = {2021},
issue_date = {Jun 2021},
publisher = {Now Publishers Inc.},
address = {Hanover, MA, USA},
volume = {14},
number = {1--2},
issn = {1935-8237},
url = {https://doi.org/10.1561/2200000083},
doi = {10.1561/2200000083},
journal = {Found. Trends Mach. Learn.},
month = jun,
pages = {1--210},
numpages = {214}
}

@inproceedings{
yu2018slimmable,
title={Slimmable Neural Networks},
author={Jiahui Yu and Linjie Yang and Ning Xu and Jianchao Yang and Thomas Huang},
booktitle={International Conference on Learning Representations},
year={2019},
url={https://openreview.net/forum?id=H1gMCsAqY7},
}

@article{Yu2019UniversallySN,
  title={Universally Slimmable Networks and Improved Training Techniques},
  author={Jiahui Yu and Thomas S. Huang},
  journal={2019 IEEE/CVF International Conference on Computer Vision (ICCV)},
  year={2019},
  pages={1803-1811},
  url={https://api.semanticscholar.org/CorpusID:76660361}
}

@inproceedings{
  cai2020once,
  title={Once for All: Train One Network and Specialize it for Efficient Deployment},
  author={Han Cai and Chuang Gan and Tianzhe Wang and Zhekai Zhang and Song Han},
  booktitle={International Conference on Learning Representations},
  year={2020},
  url={https://arxiv.org/pdf/1908.09791.pdf}
}

@InProceedings{pmlr-v162-wortsman22a,
  title = 	 {Model soups: averaging weights of multiple fine-tuned models improves accuracy without increasing inference time},
  author =       {Wortsman, Mitchell and Ilharco, Gabriel and Gadre, Samir Ya and Roelofs, Rebecca and Gontijo-Lopes, Raphael and Morcos, Ari S and Namkoong, Hongseok and Farhadi, Ali and Carmon, Yair and Kornblith, Simon and Schmidt, Ludwig},
  booktitle = 	 {Proceedings of the 39th International Conference on Machine Learning},
  pages = 	 {23965--23998},
  year = 	 {2022},
  editor = 	 {Chaudhuri, Kamalika and Jegelka, Stefanie and Song, Le and Szepesvari, Csaba and Niu, Gang and Sabato, Sivan},
  volume = 	 {162},
  series = 	 {Proceedings of Machine Learning Research},
  month = 	 {17--23 Jul},
  publisher =    {PMLR},
  url = 	 {https://proceedings.mlr.press/v162/wortsman22a.html}
}

@inproceedings{
ilharco2023editing,
title={Editing models with task arithmetic},
author={Gabriel Ilharco and Marco Tulio Ribeiro and Mitchell Wortsman and Ludwig Schmidt and Hannaneh Hajishirzi and Ali Farhadi},
booktitle={The Eleventh International Conference on Learning Representations },
year={2023},
url={https://openreview.net/forum?id=6t0Kwf8-jrj}
}

@inproceedings{
matena2022merging,
title={Merging Models with Fisher-Weighted Averaging},
author={Michael S Matena and Colin Raffel},
booktitle={Advances in Neural Information Processing Systems},
editor={Alice H. Oh and Alekh Agarwal and Danielle Belgrave and Kyunghyun Cho},
year={2022},
url={https://openreview.net/forum?id=LSKlp_aceOC}
}

@misc{qwen2025qwen25technicalreport,
      title={Qwen2.5 Technical Report}, 
      author={{Qwen Team} and An Yang and Baosong Yang and Beichen Zhang and Binyuan Hui and Bo Zheng and Bowen Yu and Chengyuan Li and Dayiheng Liu and Fei Huang and Haoran Wei and Huan Lin and Jian Yang and Jianhong Tu and Jianwei Zhang and Jianxin Yang and Jiaxi Yang and Jingren Zhou and Junyang Lin and Kai Dang and Keming Lu and Keqin Bao and Kexin Yang and Le Yu and Mei Li and Mingfeng Xue and Pei Zhang and Qin Zhu and Rui Men and Runji Lin and Tianhao Li and Tianyi Tang and Tingyu Xia and Xingzhang Ren and Xuancheng Ren and Yang Fan and Yang Su and Yichang Zhang and Yu Wan and Yuqiong Liu and Zeyu Cui and Zhenru Zhang and Zihan Qiu},
      year={2025},
      eprint={2412.15115},
      archivePrefix={arXiv},
      primaryClass={cs.CL},
      url={https://arxiv.org/abs/2412.15115}, 
}

@misc{alpaca,
  author = {Rohan Taori and Ishaan Gulrajani and Tianyi Zhang and Yann Dubois and Xuechen Li and Carlos Guestrin and Percy Liang and Tatsunori B. Hashimoto },
  title = {Stanford Alpaca: An Instruction-following LLaMA model},
  year = {2023},
  publisher = {GitHub},
  journal = {GitHub repository},
  howpublished = {\url{https://github.com/tatsu-lab/stanford_alpaca}},
}

@misc{twenty_newsgroups_113,
  author       = {Mitchell, Tom},
  title        = {{Twenty Newsgroups}},
  year         = {1997},
  howpublished = {UCI Machine Learning Repository},
  note         = {{DOI}: https://doi.org/10.24432/C5C323}
}

@inproceedings{
tastan2026loft,
title={Lo{FT}: Low-Rank Adaptation That Behaves Like Full Fine-Tuning},
author={Nurbek Tastan and Stefanos Laskaridis and Martin Tak{\'a}{\v{c}} and Karthik Nandakumar and Samuel Horv{\'a}th},
booktitle={The Fourteenth International Conference on Learning Representations},
year={2026},
url={https://openreview.net/forum?id=86P3sb1dpr}
}

@inproceedings{
horvath2021,
title={Fj{ORD}: Fair and Accurate Federated Learning under heterogeneous targets with Ordered Dropout},
author={Samuel Horv{\'a}th and Stefanos Laskaridis and Mario Almeida and Ilias Leontiadis and Stylianos Venieris and Nicholas Donald Lane},
booktitle={Advances in Neural Information Processing Systems},
editor={A. Beygelzimer and Y. Dauphin and P. Liang and J. Wortman Vaughan},
year={2021},
url={https://openreview.net/forum?id=4fLr7H5D_eT}
}

@inproceedings{
mei2022,
title={Resource-Adaptive Federated Learning with All-In-One Neural Composition},
author={Yiqun Mei and Pengfei Guo and Mo Zhou and Vishal Patel},
booktitle={Advances in Neural Information Processing Systems},
editor={Alice H. Oh and Alekh Agarwal and Danielle Belgrave and Kyunghyun Cho},
year={2022},
url={https://openreview.net/forum?id=wfel7CjOYk}
}

@misc{
yu2020,
title={AutoSlim: Towards One-Shot Architecture Search for Channel Numbers},
author={Jiahui Yu and Thomas Huang},
year={2020},
url={https://openreview.net/forum?id=H1gz_nNYDS}
}

@misc{
wang2022,
title={ProgFed: Effective, Communication, and Computation Efficient Federated Learning by Progressive Training},
author={Hui-Po Wang and Sebastian U Stich and Yang He and Mario Fritz},
year={2022},
url={https://openreview.net/forum?id=Gpp1dfvZYYH}
}

@inproceedings{
tastan2025aequa,
title={Aequa: Fair Model Rewards in Collaborative Learning via Slimmable Networks},
author={Nurbek Tastan and Samuel Horv{\'a}th and Karthik Nandakumar},
booktitle={Forty-second International Conference on Machine Learning},
year={2025},
url={https://openreview.net/forum?id=Tw81RElDpe}
}

@InProceedings{tastan2025framework,
    author    = {Tastan, Nurbek and Nandakumar, Karthik},
    title     = {A Framework for Double-Blind Federated Adaptation of Foundation Models},
    booktitle = {Proceedings of the IEEE/CVF International Conference on Computer Vision (ICCV)},
    month     = {October},
    year      = {2025},
    pages     = {923-933}
}


\newpage

\appendix

\section{Algorithm}

\begin{algorithm}[H]
\caption{PreLort: Nested Local Training}
\label{alg:nested_training}
\begin{algorithmic}[1]
\Require Client $k$ with rank $r_k$, local dataset $\mathcal{D}_k$, global rank set $\mathcal{R} = \{r_1 < r_2 < \cdots < r_L\}$, adapter parameters $(\mathbf{A}_k, \mathbf{B}_k) \in \mathbb{R}^{r_k \times n} \times \mathbb{R}^{m \times r_k}$, learning rate $\eta$, local epochs $E$
\Ensure Updated adapter parameters $(\mathbf{A}_k, \mathbf{B}_k)$, pseudo-gradients $(\nabla \mathbf{A}_k, \nabla \mathbf{B}_k)$
\State $\mathcal{R}_k \gets \{r \in \mathcal{R} \mid r \leq r_k\}$ \Comment{Active training ranks for client $k$}
\For{$e = 1$ to $E$}
    \For{each minibatch $\mathcal{B} \subseteq \mathcal{D}_k$}
        \State $\mathcal{L}_k \gets 0$
        \For{each $r \in \mathcal{R}_k$}
            \State Construct truncated adapter: $\mathbf{A}_k^{(r)} \gets \mathbf{A}_k[{:r}, :]$,\; $\mathbf{B}_k^{(r)} \gets \mathbf{B}_k[:, {:r}]$
            \State $\mathcal{L}_k \gets \mathcal{L}_k + \mathcal{L}_{\mathrm{CE}}\!\left(f(\cdot;\, \theta,\, r),\, \mathcal{B}\right)$
        \EndFor
        \State $\mathcal{L}_k \gets \tfrac{1}{|\mathcal{R}_k|}\,\mathcal{L}_k$ \Comment{Average over sub-rank losses}
        \State $(\mathbf{A}_k, \mathbf{B}_k) \gets (\mathbf{A}_k, \mathbf{B}_k) - \eta\,\nabla_{(\mathbf{A}_k, \mathbf{B}_k)}\,\mathcal{L}_k$
    \EndFor
\EndFor
\State Compute pseudo-gradients: $\nabla \mathbf{A}_k \gets \mathbf{A}_k^{(0)} - \mathbf{A}_k$,\; $\nabla \mathbf{B}_k \gets \mathbf{B}_k^{(0)} - \mathbf{B}_k$
\State \Return $(\nabla \mathbf{A}_k, \nabla \mathbf{B}_k)$
\end{algorithmic}
\end{algorithm}

\begin{algorithm}[H]
\caption{PreLort: Segment-wise Aggregation}
\label{alg:segmentwise_agg}
\begin{algorithmic}[1]
\Require Pseudo-gradients $\{(\nabla \mathbf{A}_k, \nabla \mathbf{B}_k)\}_{k \in \mathcal{K}}$ from participating clients, client ranks $\{r_k\}_{k \in \mathcal{K}}$, dataset sizes $\{n_k\}_{k \in \mathcal{K}}$, rank levels $\mathcal{R} = \{r_1 < r_2 < \cdots < r_L\}$ with $r_0 = 0$, global adapter $(\mathbf{A}^{\mathrm{global}}, \mathbf{B}^{\mathrm{global}})$
\Ensure Updated global adapter $(\mathbf{A}^{\mathrm{global}}, \mathbf{B}^{\mathrm{global}})$
\For{$l = 1$ to $L$}
    \State $\mathcal{K}_l \gets \{k \in \mathcal{K} \mid r_k \geq r_l\}$ \Comment{Clients contributing to segment $l$}
    \State $N_l \gets \sum_{k \in \mathcal{K}_l} n_k$
    \State $\nabla \mathbf{A}^{\mathrm{global}}_{r_{l-1}:r_l} \gets \dfrac{1}{N_l} \sum_{k \in \mathcal{K}_l} n_k\, \nabla \mathbf{A}_k[r_{l-1}:r_l,\, :]$
    \State $\nabla \mathbf{B}^{\mathrm{global}}_{r_{l-1}:r_l} \gets \dfrac{1}{N_l} \sum_{k \in \mathcal{K}_l} n_k\, \nabla \mathbf{B}_k[:,\, r_{l-1}:r_l]$
\EndFor
\State Assemble $\nabla \mathbf{A}^{\mathrm{global}}$ and $\nabla \mathbf{B}^{\mathrm{global}}$ from all segments
\State $\mathbf{A}^{\mathrm{global}} \gets \mathbf{A}^{\mathrm{global}} + \nabla \mathbf{A}^{\mathrm{global}}$
\State $\mathbf{B}^{\mathrm{global}} \gets \mathbf{B}^{\mathrm{global}} + \nabla \mathbf{B}^{\mathrm{global}}$
\State \Return $(\mathbf{A}^{\mathrm{global}}, \mathbf{B}^{\mathrm{global}})$
\end{algorithmic}
\end{algorithm}

\section{Environments, Datasets, and Metric}

\textbf{Computer Resources.} All experiments were run on a single NVIDIA RTX A6000 GPU. Additional GPUs were used to run experiments in parallel.

\textbf{Dolly dataset.} The Dolly dataset is an open-source dataset containing 15k text samples generated by Databricks employees. Topics covered include brainstorming, classification, closed QA, generation, information extraction, open QA, and summarization.

\textbf{Alpaca dataset.} The Alpaca dataset contains 52K instruction-following data samples used for fine-tuning language models. The dataset is designed to be diverse enough for fine-tuning LLMs.

\textbf{20 Newsgroup dataset.} The 20 Newsgroup dataset is a widely used benchmark collection of approximately 20,000 news documents partitioned across 20 different categories. It is commonly used for text classification tasks.

\textbf{Evaluation Metrics.} We evaluate model performance using three metrics:
\begin{itemize}[itemsep=1pt]
    \item \textbf{Accuracy:} Each model is fine-tuned on the Dolly, Alpaca, and 20 Newsgroup datasets respectively, and evaluated on the corresponding held-out test set.
    \item \textbf{ROUGE-L:} Measured on the held-out test sets for the Dolly and Alpaca datasets to assess the quality of generated text.
    \item \textbf{Perplexity:} Also computed on the held-out test sets for Dolly and Alpaca to evaluate language modeling performance.
\end{itemize}

\section{Hyperparameter Details}
In all our experiments, the batch size is 128 and the micro batch size is 16. We explored 
various learning rates ($5\times10^{-3}$, $1\times10^{-4}$, $3\times10^{-4}$, 
$1\times10^{-5}$, $5\times10^{-5}$) and found that $3\times10^{-4}$ yielded the best overall performance. 
Table~\ref{tab:rounds_epochs} summarizes the communication rounds and local epochs used 
for each experiment.

\begin{table}[h]
\centering
\caption{Communication rounds and local epochs per experiment. Rounds denotes the number 
of federated communication rounds, and Epochs denotes the number of local fine-tuning 
epochs per round.}
\label{tab:rounds_epochs}
\begin{tabular}{c|c|c|c}
\hline
\textbf{Foundation} & \textbf{Dataset} & \textbf{Rounds} & \textbf{Epochs} \\
\hline
\multirow{3}{*}{\textbf{TinyLlama}} 
    & Dolly        & 10 & 3 \\
    & Alpaca       & 10 & 3 \\
    & 20 Newsgroup & 15 & 1 \\
\hline
\multirow{3}{*}{\textbf{Qwen2.5-0.5B}} 
    & Dolly        & 10 & 3 \\
    & Alpaca       & 10 & 3 \\
    & 20 Newsgroup & 15 & 1 \\
\hline
\end{tabular}
\end{table}

\section{Perplexity Convergence}

\begin{figure}[h]
\centering
\begin{tikzpicture}

\begin{axis}[
    name=ax1,
    width=0.38\textwidth,
    height=0.30\textwidth,
    title={\small TinyLlama, Dolly},
    xlabel={\small Round},
    ylabel={\small Perplexity},
    xmin=1, xmax=10,
    ymin=4.3, ymax=7.0,
    xtick={1,3,5,7,9},
    enlargelimits=false,
    grid=major,
    grid style={dashed, gray!30},
    tick label style={font=\tiny},
    label style={font=\small},
    title style={font=\small},
]
\addplot[color=cyan!70!black, line width=1pt] coordinates {
    (1,6.523042906138913)(2,5.484782907451977)(3,4.983699653177253)(4,4.798244200827821)
    (5,4.719043200990134)(6,4.672156831417193)(7,4.6414252435650685)(8,4.6196191047150315)
    (9,4.603234112649258)(10,4.59024021949037)
};
\addplot[color=purple, line width=1pt] coordinates {
    (1,6.693176518066747)(2,5.723143970637777)(3,5.103739719186843)(4,4.839129458221711)
    (5,4.722113735272503)(6,4.660568611524965)(7,4.622434377905881)(8,4.595308864694097)
    (9,4.57484740826439)(10,4.5589721910356795)
};
\addplot[color=green!60!black, line width=1pt] coordinates {
    (1,4.699673351538286)(2,4.554852354036281)(3,4.522108900429071)(4,4.500648603159538)
    (5,4.4854041417997825)(6,4.473063504450821)(7,4.463099271102045)(8,4.454649934952109)
    (9,4.447097577732689)(10,4.440774072726686)
};
\addplot[color=orange, line width=1pt] coordinates {
    (1,4.5685244563896354)(2,4.701526247688573)(3,4.478380442573122)(4,4.45929370095344)
    (5,4.476627721521591)(6,4.517113501777359)(7,4.584858513685334)(8,4.666754318737619)
    (9,4.776142962181635)(10,4.911517652332017)
};
\addplot[color=red, line width=1pt] coordinates {
    (1,6.523758970121692)(2,5.524814513702702)(3,5.113869353581363)(4,4.922985606044373)
    (5,4.614740734951381)(6,4.596962776111497)(7,4.581971388853217)
    (8,4.5693256060539476)(9,4.558618202722524)(10,4.549175742549824)
};
\addplot[color=blue, line width=1pt] coordinates {
    (1,4.763666965276554)(2,4.546902124144931)(3,4.456233775695579)(4,4.4236613717558795)
    (5,4.407972782036106)(6,4.40132302494228)(7,4.395800683166874)(8,4.398201182851344)
    (9,4.398609706583656)(10,4.4030234364008045)
};
\addplot[draw=none] coordinates {(1,0)(10,0)};
\end{axis}

\begin{axis}[
    name=ax2,
    at={(ax1.north east)}, anchor=north west,
    xshift=0.6cm,
    width=0.38\textwidth,
    height=0.30\textwidth,
    title={\small TinyLlama, Alpaca},
    xlabel={\small Round},
    xmin=1, xmax=10,
    ymin=2.6, ymax=5.0,
    xtick={1,3,5,7,9},
    enlargelimits=false,
    grid=major,
    grid style={dashed, gray!30},
    tick label style={font=\tiny},
    label style={font=\small},
    title style={font=\small},
]
\addplot[color=cyan!70!black, line width=1pt] coordinates {
    (1,4.543686391577299)(2,3.418260130563011)(3,2.911099823262093)(4,2.7754994636274075)
    (5,2.733763598210929)(6,2.713409169518595)(7,2.701180019295749)(8,2.6929403050209824)
    (9,2.686886899020292)(10,2.682187827969596)
};
\addplot[color=purple, line width=1pt] coordinates {
    (1,4.891845562206198)(2,3.8046871111209013)(3,3.161873937023886)(4,2.8545471145471426)
    (5,2.7570470220513057)(6,2.720603151602382)(7,2.7009958890716317)(8,2.6881524039970057)
    (9,2.6790053054794276)(10,2.6721648068896666)
};
\addplot[color=green!60!black, line width=1pt] coordinates {
    (1,2.6948907027274855)(2,2.6645622029421014)(3,2.6530770437625333)(4,2.64539523916579)
    (5,2.6396434362655117)(6,2.6350428238367347)(7,2.631226223288615)(8,2.6280285852890146)
    (9,2.625352806735212)(10,2.623001550555428)
};
\addplot[color=orange, line width=1pt] coordinates {
    (1,2.664891292993552)(2,3.254698267193999)(3,2.6978730221238245)(4,2.6992407552428883)
    (5,2.7254690201176146)(6,2.768980128155323)(7,2.8261305788638227)(8,2.913363280831712)
    (9,3.004984261378897)(10,3.108354725318788)
};
\addplot[color=red, line width=1pt] coordinates {
    (1,4.05194476511584)(2,3.096940924875825)(3,2.862354166760584)(4,2.7851177988944555)
    (5,2.6861883555752555)(6,2.6817944949808714)(7,2.678128595498577)(8,2.675050789272113)
    (9,2.672284562287756)(10,2.669961678279241)
};
\addplot[color=blue, line width=1pt] coordinates {
    (1,2.9108974327397474)(2,2.6817293902045654)(3,2.6438753842490175)(4,2.6321077350678572)
    (5,2.6295233496215937)(6,2.6276455734725643)(7,2.6285553666566885)(8,2.6297402917218893)
    (9,2.633370615751865)(10,2.637903560708238)
};
\end{axis}

\begin{axis}[
    name=ax3,
    at={(ax2.north east)}, anchor=north west,
    xshift=0.6cm,
    width=0.38\textwidth,
    height=0.30\textwidth,
    title={\small TinyLlama, 20 Newsgroup},
    xlabel={\small Round},
    xmin=1, xmax=15,
    ymin=6.4, ymax=15.5,
    xtick={1,3,5,7,9,11,13,15},
    enlargelimits=false,
    grid=major,
    grid style={dashed, gray!30},
    tick label style={font=\tiny},
    label style={font=\small},
    title style={font=\small},
]
\addplot[color=cyan!70!black, line width=1pt] coordinates {
    (1,15.142078232320477)(2,13.515613453515483)(3,12.257383407167836)(4,11.245372672749589)
    (5,10.263508026262977)(6,9.322207181735068)(7,8.580133848316917)(8,8.109538233186791)
    (9,7.816438644696092)(10,7.61085533392927)(11,7.4545375464823636)(12,7.343642056843016)
    (13,7.256970433756611)(14,7.184951157194944)(15,7.1240551153139)
};
\addplot[color=purple, line width=1pt] coordinates {
    (1,14.52615142894205)(2,12.621478144110355)(3,11.250838276443881)(4,10.027222472184976)
    (5,8.987924727180438)(6,8.295336522571253)(7,7.889030700274656)(8,7.629978056974459)
    (9,7.448128079039653)(10,7.3069912702371305)(11,7.20956091984332)(12,7.125183728720132)
    (13,7.060823079861438)(14,7.009591554693666)(15,6.96115806379353)
};
\addplot[color=green!60!black, line width=1pt] coordinates {
    (1,10.300993763582046)(2,7.577660832758753)(3,7.057584019158124)(4,6.869196672606786)
    (5,6.781978104847032)(6,6.724394789984141)(7,6.6871235880908335)(8,6.6595952667129374)
    (9,6.636404761296825)(10,6.617501731004431)(11,6.601350867485234)(12,6.58927148768628)
    (13,6.577184045678023)(14,6.566156837114824)(15,6.556147521198729)
};
\addplot[color=orange, line width=1pt] coordinates {
    (1,7.4499895049763065)(2,6.821458000168179)(3,6.711911391205987)(4,6.650199973051711)
    (5,6.604850751229989)(6,6.575788368390346)(7,6.557695522671799)(8,6.544443071848287)
    (9,6.534988763888026)(10,6.526026450653682)(11,6.524304464226244)(12,6.51625477912776)
    (13,6.513289973354583)(14,6.512047881530193)(15,6.509825439591368)
};
\addplot[color=red, line width=1pt] coordinates {
    (1,10.436007053083822)(2,7.991642556337871)(3,7.235221757890753)(4,6.97815149864283)
    (5,6.8583735593807225)(6,6.788190822321825)(7,6.741569825117875)(8,6.708480783722616)
    (9,6.68465575070296)(10,6.665950320231275)(11,6.650776329517178)(12,6.638889709712982)
    (13,6.62806063755163)(14,6.618841383620591)(15,6.610519754194833)
};
\addplot[color=blue, line width=1pt] coordinates {
    (1,7.707867915604515)(2,7.494144820045743)(3,7.010826335389048)(4,6.871618114587269)
    (5,6.777869529139823)(6,6.696885232723373)(7,6.613479490589662)(8,6.577002505856794)
    (9,6.534730853792898)(10,6.5085236446847645)(11,6.495567045479857)(12,6.485686419834435)
    (13,6.481425411355738)(14,6.482824050282832)(15,6.489952125663594)
};
\end{axis}

\begin{axis}[
    name=ax4,
    at={(ax1.below south west)}, anchor=north west,
    yshift=-1.0cm,
    width=0.38\textwidth,
    height=0.30\textwidth,
    title={\small Qwen2.5-0.5B, Dolly},
    xlabel={\small Round},
    ylabel={\small Perplexity},
    xmin=1, xmax=10,
    ymin=6.7, ymax=10.0,
    xtick={1,3,5,7,9},
    enlargelimits=false,
    grid=major,
    grid style={dashed, gray!30},
    tick label style={font=\tiny},
    label style={font=\small},
    title style={font=\small, yshift=0.2cm},
]
\addplot[color=cyan!70!black, line width=1pt] coordinates {
    (1,9.871698441874836)(2,8.751578180015137)(3,8.050583091431726)(4,7.56428235496565)
    (5,7.340352479485033)(6,7.21298697544142)(7,7.140074923146727)(8,7.100742079551723)
    (9,7.071026262151867)(10,7.048190248148723)
};
\addplot[color=purple, line width=1pt] coordinates {
    (1,9.820334299447538)(2,8.697888906085657)(3,8.03736178076677)(4,7.570779047789604)
    (5,7.328870075182033)(6,7.2014965313151915)(7,7.130003815892675)(8,7.086648231181634)
    (9,7.053742712624856)(10,7.027638781069494)
};
\addplot[color=green!60!black, line width=1pt] coordinates {
    (1,7.4121554447442906)(2,7.0871798570420825)(3,7.010790779342696)(4,6.926290032282738)
    (5,6.876032685052246)(6,6.852368911497158)(7,6.834018369805411)(8,6.822551101181314)
    (9,6.8110882509900375)(10,6.802493775557059)
};
\addplot[color=orange, line width=1pt] coordinates {
    (1,7.050372427468864)(2,6.967186464509606)(3,6.911322704078538)(4,6.907036081354674)
    (5,6.940513636562946)(6,7.020961993211168)(7,7.134951035024565)(8,7.29233499133198)
    (9,7.496631006692834)(10,7.692700357881855)
};
\addplot[color=red, line width=1pt] coordinates {
    (1,9.147754053329482)(2,7.983727156200853)(3,7.4136928935059405)(4,7.2193956362804)
    (5,6.97457253896531)(6,6.96098127713261)(7,6.949112675980951)(8,6.937766808559686)
    (9,6.92678347278943)(10,6.915912272673399)
};
\addplot[color=blue, line width=1pt] coordinates {
    (1,7.264098634022448)(2,7.313498026084346)(3,6.985302598813141)(4,6.91695153395196)
    (5,6.897398767660008)(6,6.863807538666495)(7,6.8356521716358145)(8,6.8222325082252615)
    (9,6.812636591475162)(10,6.81564090656462)
};
\end{axis}

\begin{axis}[
    name=ax5,
    at={(ax4.north east)}, anchor=north west,
    xshift=0.6cm,
    width=0.38\textwidth,
    height=0.30\textwidth,
    title={\small Qwen2.5-0.5B, Alpaca},
    xlabel={\small Round},
    xmin=1, xmax=10,
    ymin=2.6, ymax=5.4,
    xtick={1,3,5,7,9},
    enlargelimits=false,
    grid=major,
    grid style={dashed, gray!30},
    tick label style={font=\tiny},
    label style={font=\small},
    title style={font=\small, yshift=0.2cm},
]
\addplot[color=cyan!70!black, line width=1pt] coordinates {
    (1,5.260744891375005)(2,4.110514771513615)(3,3.4215429414361824)(4,3.122755988686641)
    (5,2.998958226793826)(6,2.930072895985308)(7,2.886843049267515)(8,2.856920614975642)
    (9,2.8354435341982147)(10,2.8185559702618703)
};
\addplot[color=purple, line width=1pt] coordinates {
    (1,5.296803908269775)(2,4.198819043678849)(3,3.4950077925581864)(4,3.1509329914438995)
    (5,3.0009043988594253)(6,2.9237148156713824)(7,2.8747114588636458)(8,2.8404306769189462)
    (9,2.8165273620775806)(10,2.798302190937662)
};
\addplot[color=green!60!black, line width=1pt] coordinates {
    (1,3.010383147092918)(2,2.7860990940347623)(3,2.7416299605751626)(4,2.7223505473652736)
    (5,2.7095893624593517)(6,2.700391911008859)(7,2.6928343515833078)(8,2.687059123816406)
    (9,2.6819060633669487)(10,2.6774757362842316)
};
\addplot[color=orange, line width=1pt] coordinates {
    (1,2.8434565505735567)(2,2.8369693923346717)(3,2.7272256263032286)(4,2.6941194448191106)
    (5,2.6727697161892614)(6,2.6604264507295414)(7,2.6552272274090787)(8,2.6565190218262846)
    (9,2.6630065817669917)(10,2.6695139203855676)
};
\addplot[color=red, line width=1pt] coordinates {
    (1,4.582036145602445)(2,3.772887953444675)(3,3.4347552118936827)(4,3.424372331472908)
    (5,3.329715609356351)(6,3.038341463415894)(7,3.25523705314171)(8,3.4096800471020967)
    (9,3.2070377205329314)(10,3.4447499424705508)
};
\addplot[color=blue, line width=1pt] coordinates {
    (1,3.0236124116411527)(2,2.7882204356278084)(3,2.702080462523954)(4,2.6760637825399574)
    (5,2.663487473129018)(6,2.653674642363222)(7,2.6466899423699437)(8,2.6432289680255927)
    (9,2.6388288104050037)(10,2.6378768356968596)
};
\end{axis}

\begin{axis}[
    name=ax6,
    at={(ax5.north east)}, anchor=north west,
    xshift=0.6cm,
    width=0.38\textwidth,
    height=0.30\textwidth,
    title={\small Qwen2.5-0.5B, 20 Newsgroup},
    xlabel={\small Round},
    xmin=1, xmax=15,
    ymin=11.0, ymax=20.0,
    xtick={1,3,5,7,9,11,13,15},
    enlargelimits=false,
    grid=major,
    grid style={dashed, gray!30},
    tick label style={font=\tiny},
    label style={font=\small},
    title style={font=\small, yshift=0.2cm},
]
\addplot[color=cyan!70!black, line width=1pt] coordinates {
    (1,19.769484078226917)(2,17.192056654661943)(3,15.684318407994866)(4,14.45118908464323)
    (5,13.432883396136559)(6,12.734930397117813)(7,12.313065685796253)(8,12.093481965212579)
    (9,11.950333908834429)(10,11.847310695808602)(11,11.770719525442452)(12,11.715345744269925)
    (13,11.671973251488161)(14,11.636656519237)(15,11.606901315799963)
};
\addplot[color=purple, line width=1pt] coordinates {
    (1,16.971216801605525)(2,14.447806208827384)(3,13.73070553010308)(4,12.72186938583293)
    (5,11.714896052704367)(6,11.703514746896263)(7,11.688041852786754)(8,11.660334329294235)
    (9,11.63542013413518)(10,11.605234160571581)(11,11.597587900436397)(12,11.584661585702685)
    (13,11.568012463456499)(14,11.55552168176093)(15,11.541255209223236)
};
\addplot[color=green!60!black, line width=1pt] coordinates {
    (1,15.299773125531953)(2,12.266601496627924)(3,11.73070553010308)(4,11.562186938583293)
    (5,11.534896052704367)(6,11.403514746896263)(7,11.398041852786754)(8,11.350334329294235)
    (9,11.33542013413518)(10,11.315234160571581)(11,11.337587900436397)(12,11.344661585702685)
    (13,11.358012463456499)(14,11.40552168176093)(15,11.405555209223236)
};
\addplot[color=orange, line width=1pt] coordinates {
    (1,13.381442105567956)(2,12.199832003831046)(3,11.790529456438021)(4,11.591099486039303)
    (5,11.480978314682313)(6,11.434788742908799)(7,11.41946266757947)(8,11.447821438235822)
    (9,11.490978144493996)(10,11.550865804472076)(11,11.600076249387817)(12,11.685694382736608)
    (13,11.766238115730392)(14,11.858463940687129)(15,11.958174780259858)
};
\addplot[color=red, line width=1pt] coordinates {
    (1,15.264421601010769)(2,12.885293958911998)(3,11.825163967017614)(4,11.630159348694178)
    (5,11.538470225386316)(6,11.48171562127682)(7,11.442476452301449)(8,11.412571889212755)
    (9,11.389471866027987)(10,11.369677076454227)(11,11.352738497139223)(12,11.338070625047193)
    (13,11.32504823299752)(14,11.313164251902752)(15,11.302307597020285)
};
\addplot[color=blue, line width=1pt] coordinates {
    (1,13.978908328529185)(2,12.16832101734118)(3,11.811716378482178)(4,11.61373974118474)
    (5,11.472144695606554)(6,11.418826489489643)(7,11.375388580276073)(8,11.341064716145546)
    (9,11.31571621485158)(10,11.293842272660363)(11,11.274588626031166)(12,11.25851168129776)
    (13,11.246330291049947)(14,11.232811862125416)(15,11.222945674640025)
};
\end{axis}

\end{tikzpicture}
\vspace{0.3cm}
\begin{tikzpicture}
\begin{axis}[
    hide axis,
    legend columns=3,
    legend style={
        font=\small,
        draw=none,
        column sep=0.4cm,
    },
]
\addplot[color=cyan!70!black, line width=1pt] coordinates {(0,0)};
\addlegendentry{ZeroPad}
\addplot[color=purple, line width=1pt] coordinates {(0,0)};
\addlegendentry{HetLoRA}
\addplot[color=green!60!black, line width=1pt] coordinates {(0,0)};
\addlegendentry{FLoRA}
\addplot[color=orange, line width=1pt] coordinates {(0,0)};
\addlegendentry{Nested Agg Only}
\addplot[color=red, line width=1pt] coordinates {(0,0)};
\addlegendentry{Nested Train Only}
\addplot[color=blue, line width=1pt] coordinates {(0,0)};
\addlegendentry{PreLort}
\end{axis}
\end{tikzpicture}
\vspace{-2.2cm}
\caption{Perplexity convergence across communication rounds for TinyLlama and Qwen2.5-0.5B on Dolly, Alpaca, and 20 Newsgroup datasets.}
\label{fig:perplexity_convergence}
\end{figure}
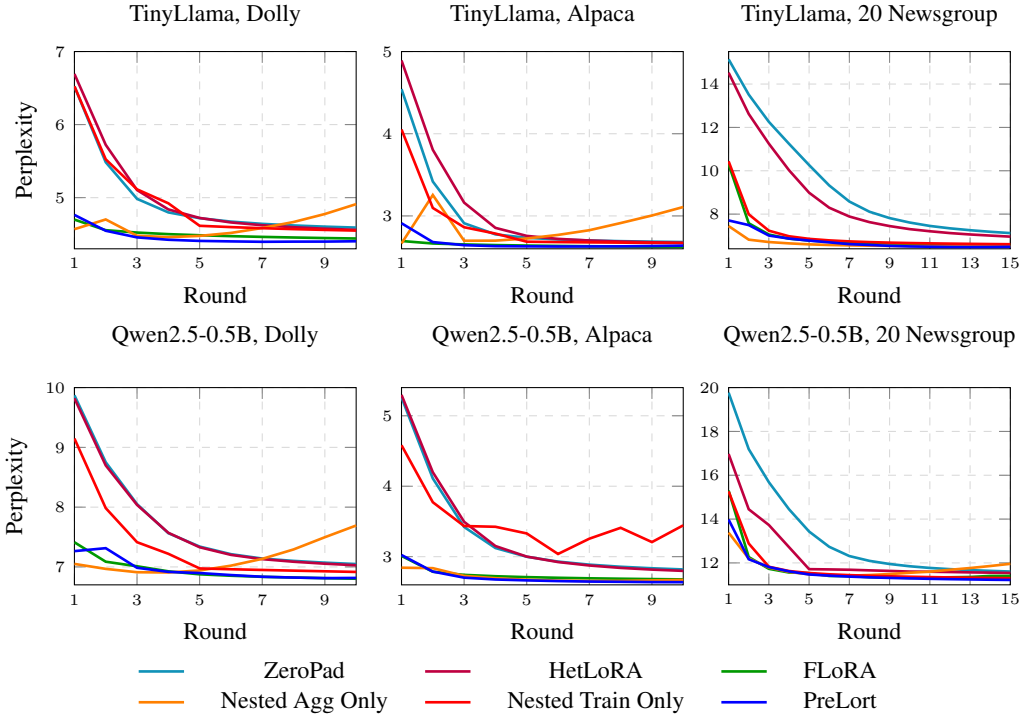

Figure~\ref{fig:perplexity_convergence} shows the perplexity convergence of all methods across communication rounds on both models and all three datasets. PreLort consistently achieves competitive or lower final perplexity compared to baselines, though the visual gap is diminished due to the high initial perplexity of ZeroPad and HetLoRA dominating the y-axis scale. Nested Aggregation Only diverges over rounds on most settings, confirming that segment-wise aggregation alone is insufficient without the corresponding local training strategy. Similarly, Nested Training Only exhibits poor or sometimes unstable convergence, suggesting that segment-wise local training must be coupled with the corresponding aggregation strategy to achieve stable performance.





\end{document}